\documentclass[12pt]{article}
\usepackage{authblk}
\usepackage{amsmath}
\usepackage{float}

\usepackage{graphicx}
\usepackage{dcolumn}
\usepackage{bm}

\usepackage[natbib=true,style=numeric,sorting=none]{biblatex}
\usepackage[right=2.5cm, bottom=2.5cm, left=2.5cm, top=2cm]{geometry}

\usepackage{xcolor}

\usepackage{lineno}

\begin{document}

\title{Characterization of SiPM Performance in a Small Satellite in Low Earth Orbit using LabOSat-01}

\author[1]{\small Lucas Finazzi\footnote{Lead author: lfinazzi@unsam.edu.ar}}
\author[1]{Federico Izraelevitch}
\author[1,2]{Mariano Barella}
\author[3]{Fernando Gomez Marlasca}
\author[4]{Gabriel Sanca\footnote{Corresponding author: gsanca@unsam.edu.ar}}
\author[1]{Federico Golmar}

\affil[1]{Instituto de Ciencias Físicas, Universidad de San Martin, CONICET, Buenos Aires, Argentina}
\affil[2]{Department of Physics, University of Fribourg, Fribourg, Switzerland}
\affil[3]{Comisión Nacional de Energía Atómica, Buenos Aires, Argentina}
\affil[4]{Escuela de Ciencia y Tecnología, Universidad de San Martin, Buenos Aires, Argentina}

\date{\today}

\maketitle

\begin{abstract}
    In this work, the performance of SensL MicroFC-60035 SiPM devices was studied during a 1460-day mission in Low Earth Orbit (LEO) using the LabOSat-01 characterization payload. Two of these platforms, carrying two SiPMs each, were integrated into the ÑuSat-7 satellite (COSPAR-ID: 2020-003B). Analysis revealed that these SiPMs experienced an increase in dark current over time due to damage from trapped and solar proton radiation. The total ionizing dose received by the payload and the SiPMs was measured using p-MOSFET dosimeters, with a resulting value of 5~Gy, or a 1~MeV neutron equivalent fluence of $\phi_n = 5 \cdot 10^9$~n/cm$^2$. The dark current was observed to increase up to 500 times. Parameters such as Gain and Photon Detection Efficiency remained unchanged throughout the mission. These findings align with previous performance reports involving different SiPMs irradiated with various particles and energies.
\end{abstract}

\maketitle 


\section{Introduction}
\label{sec:introduction}

Silicon photomultipliers (SiPMs) are advanced solid-state optoelectronic devices that offer numerous advantages over traditional photomultiplier tubes. These advantages include higher photon detection efficiency and improved temporal resolution, among others~\cite{sipm_review1, sipm_review2}. SiPMs are compact, durable, immune to magnetic fields, and require relatively low bias voltage, making them highly attractive for space applications that require particle detection. Recently, SiPMs have been used in various space applications, including measuring transient gamma rays~\cite{grid}, detecting high-energy cosmic rays~\cite{lazio}, and detecting coincident gamma-ray bursts with gravitational wave events~\cite{gecam}. Additionally, SiPMs are used in terrestrial applications, such as communications~\cite{sipm_comms}, astrophysics~\cite{auger_sipm}, and quantum optics~\cite{nature_sipm}.

The use of these sensors in space, in particular in Low Earth Orbit (LEO), requires a thorough understanding on radiation in the satellite's environment. These sensors can suffer major degradation in radiation-prone orbits. Nevertheless, SiPMs can continue to work (albeit at reduced performance) even after degradation from ionizing or non-ionizing radiation~\cite{sipm_irradiation_1, sipm_irradiation_2, sipm_irradiation_3}. SiPM damage with medium fluences ($\sim 10^{10}$~n/cm$^2$ for 1~MeV equivalent neutrons) results mainly in an increase in their Dark Count Rate and their dark current, while maintaining other intrinsic characteristics (like Gain, Photon Detection Efficiency, and others) intact~\cite{sipm_irradiation_1}.

Four SiPMs were integrated into the ÑuSat-7 satellite (COSPAR-ID 2020-003B) on January 2020 and put in orbit using two LabOSat-01 \mbox{(LS-01)} Payloads, along with dedicated daughter boards (DBs) specifically designed for this mission (See~\cite{barella2020, finazzi_dos_2024} and the references therein). The integration of various SiPMs allowed us to have redundancies in case of electronics or SiPM malfunction. We aimed to measure the degradation of these devices after almost 4 years of mission time and show that they continue to operate, albeit with reduced performance, to controlled light stimuli.


In Section~\ref{sec:labosat}, a description of the electronics of the \mbox{LS-01} platform is presented and the tests performed on the devices of interest are also detailed. In Section~\ref{sec:orbit_characteristics}, particle radiation in the satellite's orbit is discussed. In Section~\ref{sec:results}, results from 1460 days in LEO are presented. In Section~\ref{sec:conclusions}, the conclusions of the work are outlined.

\section{The LabOSat Payload}
\label{sec:labosat}

\subsection{Electronics}

\begin{figure}[!h]
\centering
   \includegraphics[width=0.65\textwidth]{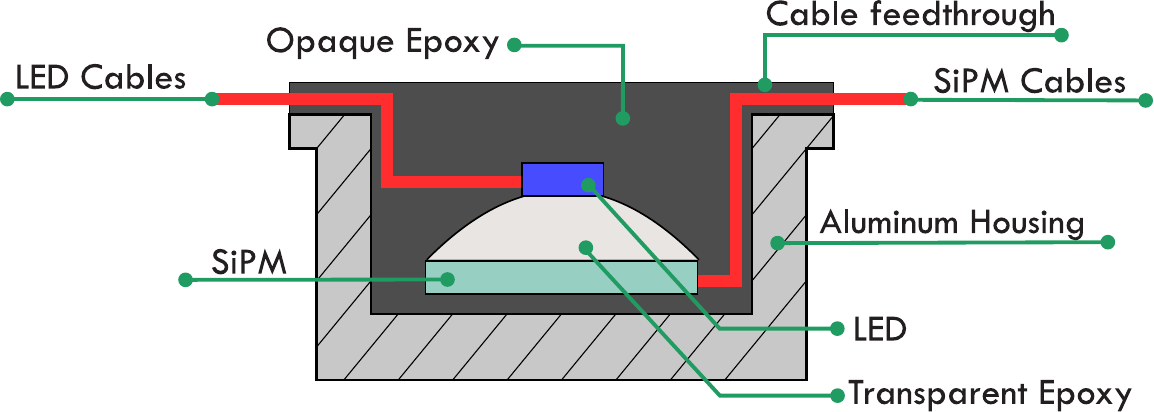}
   \caption{Schematic diagram of the SiLH design, in a cutout view (not to scale). Image taken from~\cite{barella2020}.}
   \label{fig:SiLH}
\end{figure}

The LS-01 platform and a dedicated DB were used to perform measurements on SiPMs under controlled light stimuli. SensL MicroFC-60035 SiPMs were bundled together with an LED and placed together inside a light-tight housing (SiLH, for SiPM Light-tight Housing). This allowed for the characterization of SiPMs under controlled illumination during the mission. A schematic drawing of the SiLHs, which represent the DUTs of the present study, is shown in Figure~\ref{fig:SiLH}.

The SiPM was attached to the LED using a transparent epoxy, placing the active sides of each component facing each other. The assembly was then soldered to cables, and submerged into a black opaque epoxy inside an aluminum housing. 

\begin{figure}[!h]
\centering
   \includegraphics[width=0.65\textwidth]{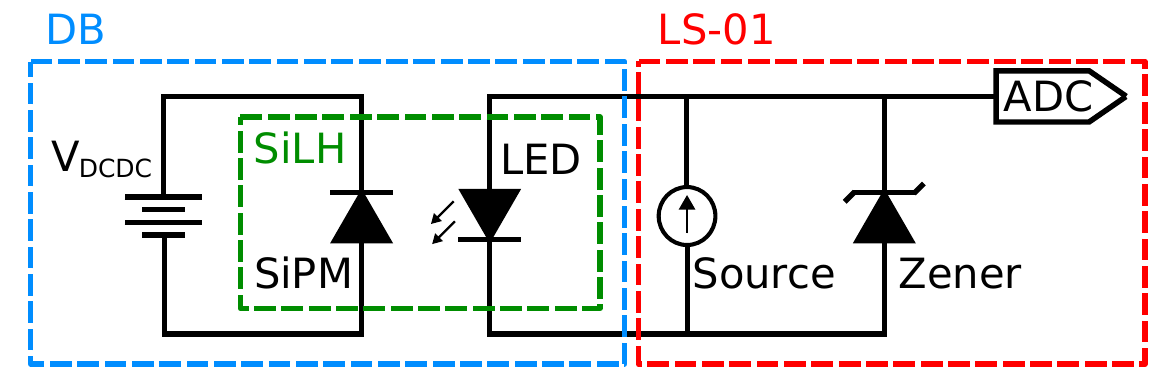}
   \caption{Depiction of DB~+~LS-01 electronics. A Zener diode was placed in the current biasing circuit to protect LS-01's ADC. Image taken from~\cite{cae_2024}.}
   \label{fig:sipm_bias_circuit}
\end{figure}

Each DB contains two SiLHs in two redundant blocks in parallel, one for each SiLH. The AD590 temperature sensor was used in the DB near the SiLHs to measure the temperature of the SiPMs. Each SiPM has its own independent power supply, based on a DC-DC boost converter chip (LT3571). The SiPMs were biased at fixed bias voltage of $(29.1 \pm 0.1)$~V. Typically, this bias value gives an overvoltage of 4.5~V at 21~ºC. The SiPM current was measured using the Monitor output pin of the DC-DC chip, which mirrors the current in the output pin (with a x5 attenuation) onto a 10~k$\Omega$ resistor. In addition, the payload contains a circuit for p-MOS dosimeter readout for a measurement on the Total Ionizing Dose (TID) on the board and SiPM sensors. All logic signals controlling LED biasing, SiPM, dosimeter and temperature readout were controlled by the microcontroller unit of LS-01. The LED and SiPM bias circuit is shown in Figure~\ref{fig:sipm_bias_circuit}. 

A more detailed description of the electronics can be found in~\cite{barella2020}.

\subsection{Test descriptions}

Two distinct tests were executed in the reported mission: The standard test and the long test. The first one was designed to test SiPM sensors under 21 LED illumination levels. These illumination levels were chosen with the following criteria: The SiPM is almost unresponsive for the first LED current values and the largest LED current value produces an illumination (henceforth called maximum LED illumination) that does not saturate the SiPM's output. For each LED current level, the LS-01 measures:

\begin{enumerate}
    \item Temperature on the DB
    \item LED voltage
    \item SiPM bias voltage
    \item SiPM current
\end{enumerate}



For more details into this test, see~\cite{barella2020}.

The long test was designed to test the SiPM sensors during a whole satellite orbit (which has a period of approximately 90~minutes). In a long test, the LED is kept off and measurements are performed once every 15 seconds for a duration of 100 minutes. Every 15 seconds, LS-01 measures

\begin{enumerate}
    \item Temperature on the DB
    \item SiPM bias voltage
    \item SiPM current
\end{enumerate}

One Payload executed the standard test throughout the mission, while the other executed a standard test from January 2020 - February 2023 and a long test from that date until January 2024.

\section{Particle Radiation and Satellite Shielding}
\label{sec:orbit_characteristics}

The ÑuSat-7 satellite was launched on January 15th 2020 (COSPAR-ID 2020-003B). The original orbit has 476~km/490~km Perigee/Apogee altitude and an inclination of 97.34 degrees. SPENVIS~\cite{spenvis} simulations were used to estimate the solar proton and the trapped electron and proton differential fluences in the aforementioned orbit for a duration of 1300 days. SAPPHIRE (total fluence) model was used for solar protons and AP-8 and AE-8 (solar minimum) trapped particle radiation models were used for trapped protons and electrons. 




The shielding provided by the satellite and by the electronics surrounding our payload was estimated to be the equivalent of 4~mm thickness aluminum~\cite{finazzi_dos_2024}. A TRIM~\cite{srim} simulation was run for this shielding and it was observed that only protons with an energy higher than $E_{min} = 28$~MeV pass through it. The trapped electron energy is too small to penetrate this equivalent shielding~\cite{e_penetration1}. This suggests that the SiPM observed damage will come mainly from protons.

It is often useful to calculate the 1~MeV neutron equivalent fluence for a differential proton distribution $\frac{d\phi}{dE}(E)$ such as the one observed in LEO orbits. To perform this calculation, the NIEL (Non-Ionizing Energy Loss) scaling hypothesis is assumed~\cite{NIEL_scaling1, NIEL_scaling2}. The 1~MeV neutron equivalent fluence $\phi_n$ can be calculated with the following equation~\cite{sipm_comp3}

\begin{equation}
    \phi_n(\mathrm{1~MeV}) = \int_{E_{\mathrm{min}}}^{E_{\mathrm{max}}} K(E) \frac{d\phi}{dE}(E) dE \ ,
\end{equation}

\noindent where $K(E)$ is the hardness factor for protons in Silicon. These factors are tabulated and can be found in many reference documents like~\cite{hardness_factor}. The 1~MeV neutron equivalent fluence for the whole mission can then be obtained from the SPENVIS and TRIM simulations and they are $\phi_n^{(\mathrm{solar})} = 1.6 \cdot 10^{10}$~n/cm$^{2}$ and $\phi_n^{(\mathrm{trapped})} = 8 \cdot 10^9$~n/cm$^{2}$, which gives $\phi_n^{(\mathrm{total})} = 2.4 \cdot 10^{10}$~n/cm$^{2}$.

The mission TID was measured to be approximately $D = 5$~Gy. To convert this value to 1~MeV neutron equivalent fluence, the following relationship was used~\cite{dose_to_neq}

\begin{equation}
    \phi_n = R_p D \ ,
\end{equation}

\noindent where $R_p$ is the neutron-proton Damage Equivalence Factor~\cite{dose_to_neq}. This factor is roughly constant in Silicon for a wide range of energies (20~MeV to 200~MeV) and has a value of $R_p \simeq 10^9$~cm$^{-2}$Gy$^{-1}$. This means that the measured 1~MeV neutron equivalent fluence is $\phi_n = 5 \cdot 10^9$~n/cm$^2$ in 1460 days or $\phi_n = 3.4 \cdot 10^6$~n/cm$^2$ per day on average. While it is approximately of the same order, the measured value is 4 times smaller than the previously estimated total fluence value using SPENVIS. The discrepancy could be due to uncertainty in the estimated thickness of the equivalent shielding (and thus, uncertainty for the estimated value of $E_{min}$). It could also be due to the fact that the dosimeter's aluminum packaging, which was not considered in the SPENVIS calculation, provides more effective shielding.




\section{Results} 
\label{sec:results}

\subsection{Standard Test}

\begin{figure}[!h]
\centering
    \includegraphics[width=0.8\textwidth]{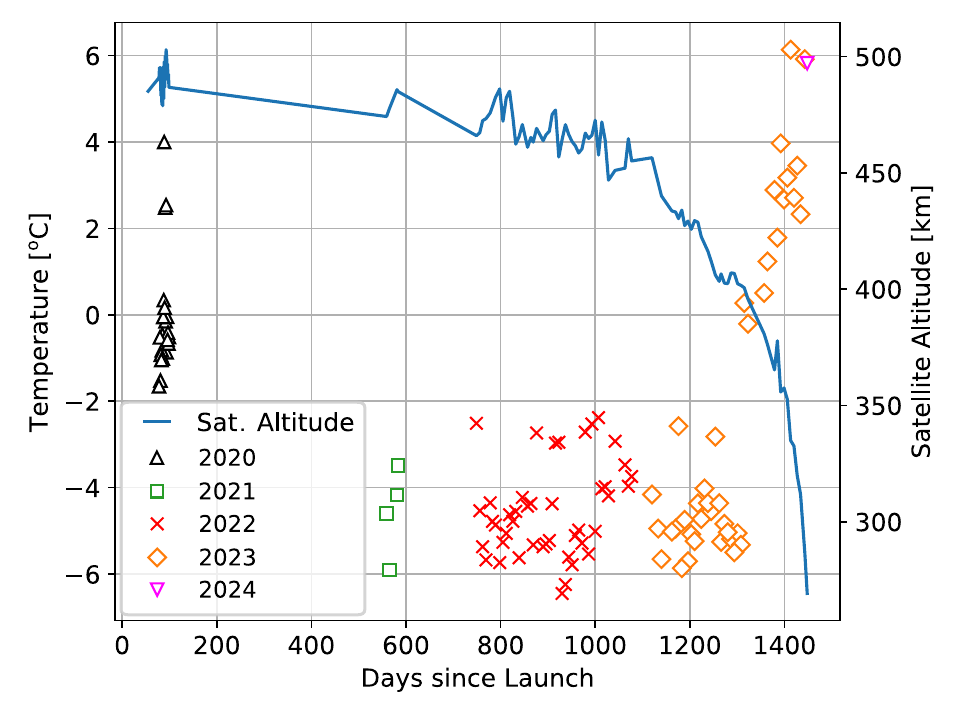}
    \caption{DB temperature of one payload vs. days since launch for the whole mission duration. Satellite altitude is also shown as a solid line. It can be observed that 2020 and the last datapoints of 2023 have a higher temperature than the other data points. The increase in temperature at the end of the mission is due to the decreasing altitude of the satellite at the end of its lifespan.}
    \label{fig:temp_vs_dsl}
\end{figure}

The first parameter that was measured in each standard test was the DB temperature, which is close to the SiPM sensors. These measurements can be seen vs. days since launch in Figure~\ref{fig:temp_vs_dsl}. The last datapoints of 2023 show an increasing temperature tendency when compared to datapoints in the previous years. The increase in temperature is due to the decreasing altitude of the satellite during the end of its lifespan. The vast majority of datapoints are contained in a temperature range of $(-3 \pm 3)$~$^{\mathrm{o}}$C. 

\begin{figure}[!h]
\centering
    \includegraphics[width=0.7\textwidth]{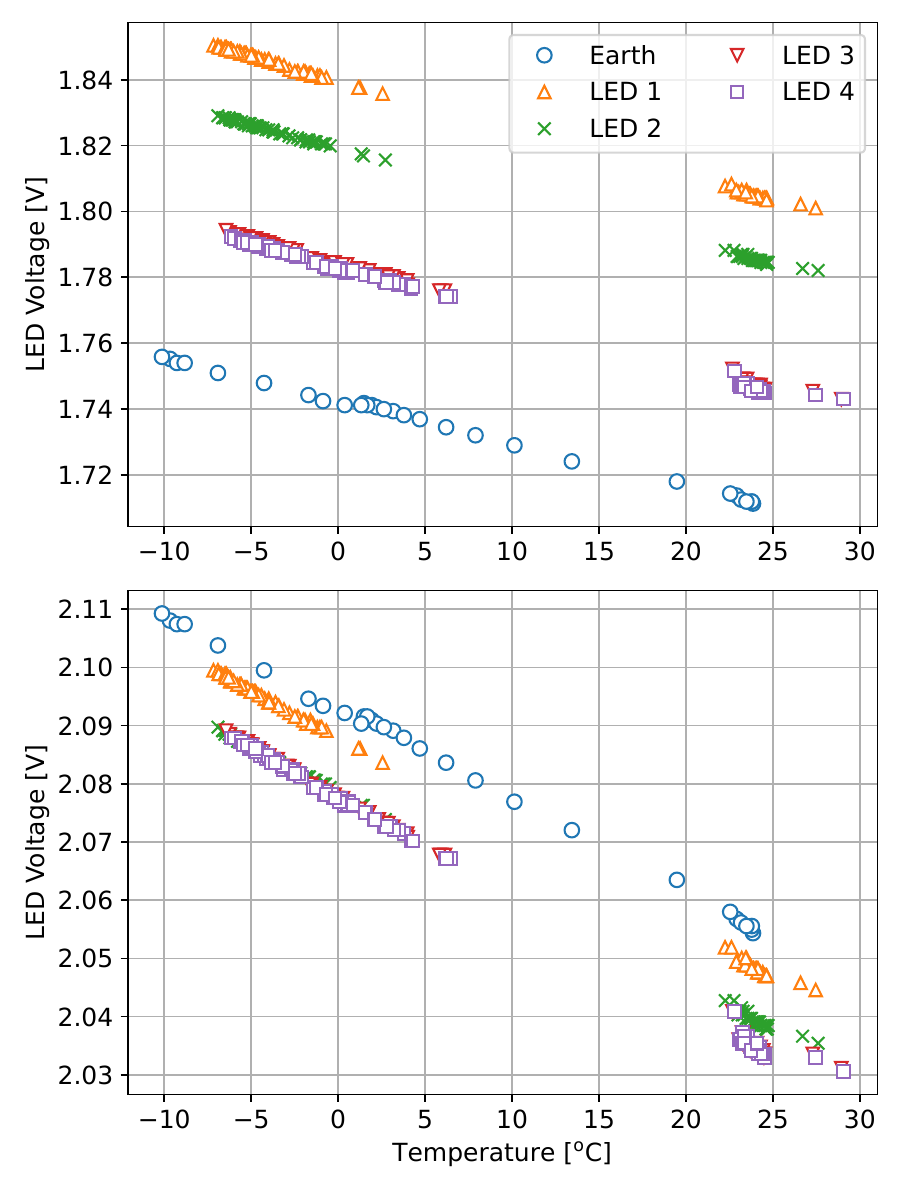}
    \caption{\textbf{Top plot:} LED voltage vs. temperature for the minimum LED current (LED off). \textbf{Bottom plot:} Maximum LED illumination (below) for all 4 LEDs in each SiLH. The Earth LED measurements are from a representative LED which was housed in a twin board and was measured on Earth after the Payloads were in orbit. LEO and Earth measurements were compared to validate the correct operation of the LEDs inside the satellite. It can be observed that the behaviour of the LEDs remains unchanged when compared to Earth measurements.}
    \label{fig:led_operation}
\end{figure}

Secondly, LED voltage was measured to monitor deviations from typical operation (for example, due to radiation damage). Figure~\ref{fig:led_operation} shows LED voltage as a function of temperature for the four SiLH components. A linear trend is observed for each device, which is compatible to Earth measurements. In addition, the slope observed for each operation point of the LEDs remained unchanged. This seems to indicate that there was no noticeable damage to the LEDs.

\begin{figure}[!h]
\centering
    \includegraphics[width=0.8\textwidth]{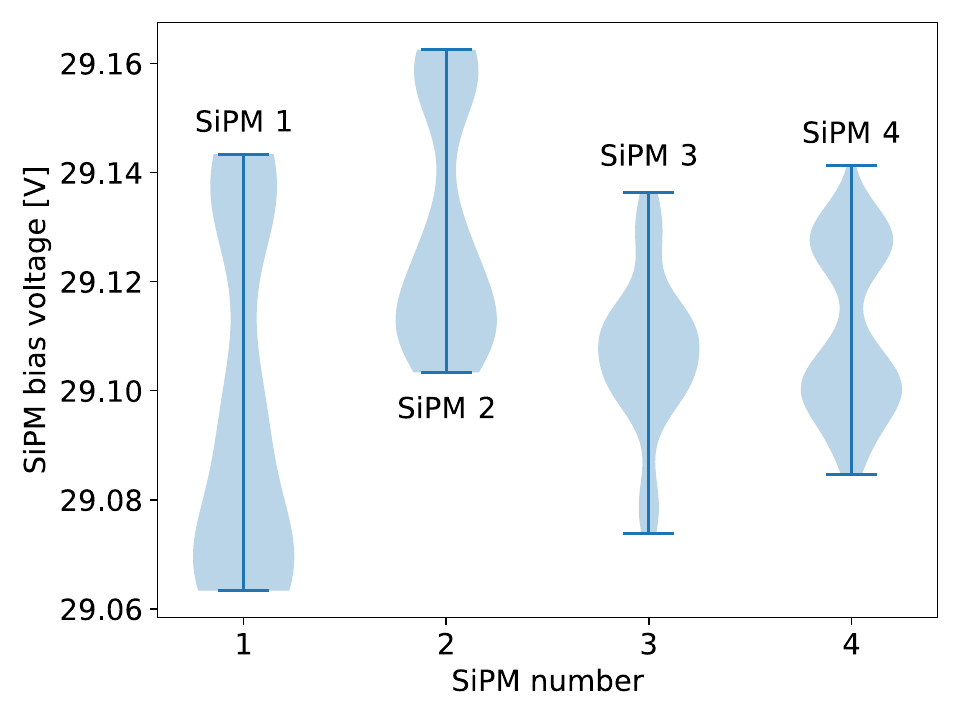}
    \caption{Violin plot of SiPM bias voltage for all 4 SiPMs used in every payload. The SiPM bias voltage for all devices never changed from the set value of $(29.1 \pm 0.1)$~V. This is a good validation for the use of LT3571 for space applications.} 
    \label{fig:dcdc_hist}
\end{figure}

The next parameter studied was the SiPM bias voltage. This value is measured during every experiment for each SiPM and all these values are shown on a violin plot in Figure~\ref{fig:dcdc_hist}. The set bias for all SiPMs was $(29.1 \pm 0.1)$~V and it can be observed that this value didn't change for 4~years of mission time. The LT3571 was found to be insensitive to temperature fluctuations and to radiation damage, making it ideal for space applications inside small satellites. The dose received by the electronics was measured to be approximately 5~Gy~\cite{finazzi_dos_2024}.

\begin{figure}[!h]
\centering
    \includegraphics[width=0.7\textwidth]{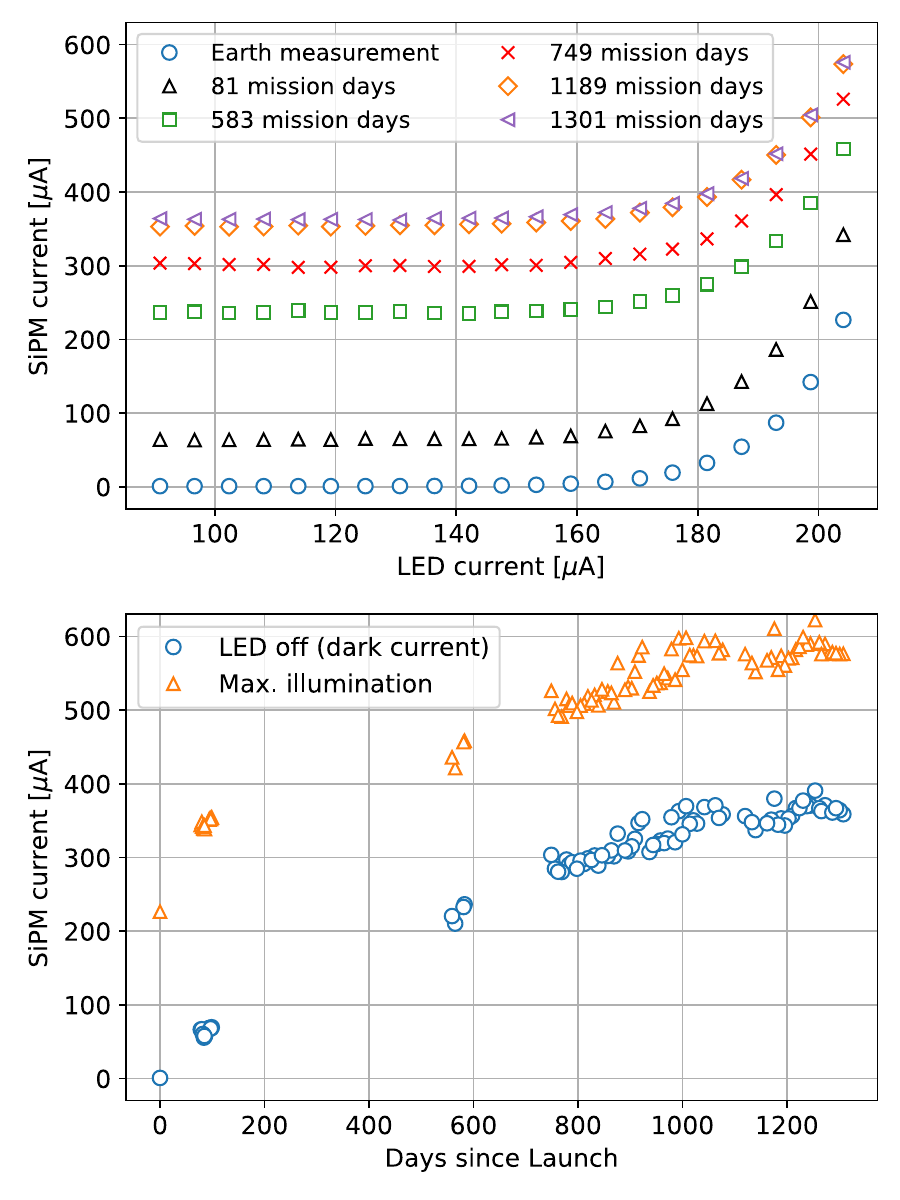}
    \caption{\textbf{Top plot:} SiPM current vs. LED current for different mission days and their comparison to Earth measurements for a particular SiPM in orbit. A clear rise in dark current can be seen and is due to radiation damage in the satellite's orbit. \textbf{Bottom plot:} SiPM current vs. days since launch for LED off (dark current) and LED at maximum illumination. The increase in dark current can clearly be seen again in this plot. What is more interesting is that the SiPM responsivity remains almost unchanged even after 4~years of mission time. For the same amount of LED current, the photocurrent generated by the SiPMs over the dark counts remained almost constant during the whole mission.}
    \label{fig:ii_and_perf}
\end{figure}

Finally, SiPM current was measured for several incident light intensities. Plots of SiPM current as a function of LED current and as a function of days since launch can be seen in Figure~\ref{fig:ii_and_perf} for one particular SiPM in orbit. From now on, results presented are limited to this particular SiPM. The behaviour of the other three was observed to be similar. For all the datapoints in these figures, only measurements in the temperature range of $(-3 \pm 3)$~$^{\mathrm{o}}$C were considered, so as to avoid temperature dependent effects (like dark current increase) when comparing measurements performed during the mission lifespan.

A clear increase in SiPM dark current can be seen, which can be attributed to radiation damage caused by protons. As observed in the bottom plot of Figure~\ref{fig:ii_and_perf}, the increase in dark current is steeper at the beginning and becomes less pronounced with time. In addition, if we observe the difference between the dark current (LED off) and maximum illumination curves, we can see that it remains unchanged for the entire duration of the mission. This may indicate that, while the dark current increased with increasing time, the SiPM Gain remained almost constant for the entire mission. It is a known fact that SiPM Gain and Photon Detection Efficiency is not degraded by medium-fluence irradiations~\cite{sipm_irradiation_1}, which is the case of this work. 

\begin{figure}[!h]
\centering
    \includegraphics[width=0.7\textwidth]{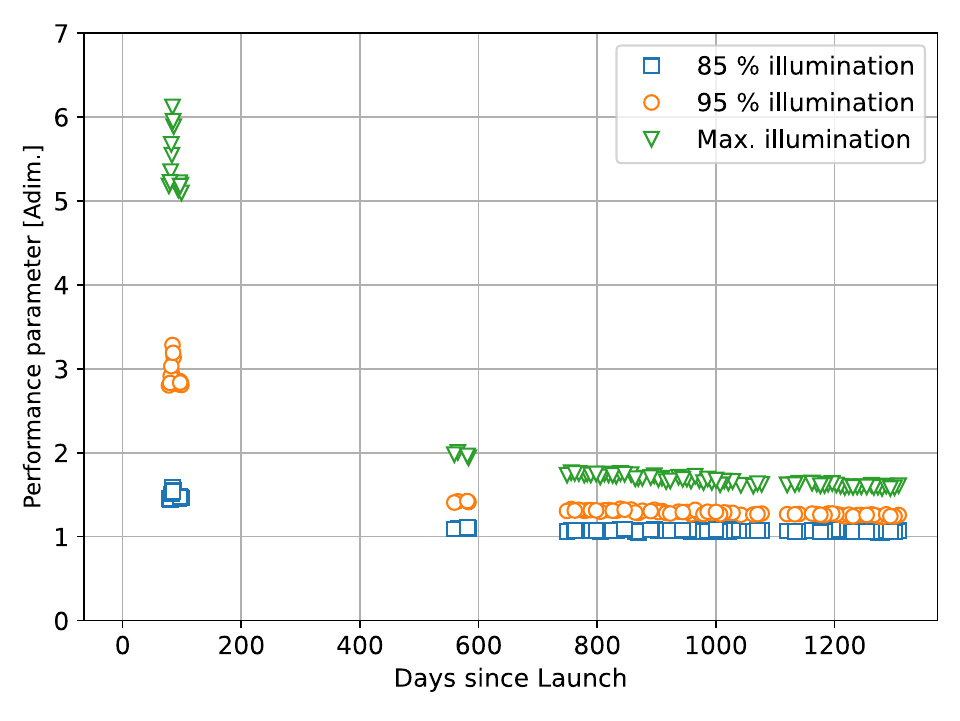}
    \caption{SiPM performance parameter vs. days since launch for different LED illuminations. The performance parameter is the ratio between SiPM current at a given LED illumination and the SiPM dark current. It can be seen that lowest illumination currents on the LED result in a current performance parameter of one, which means the SiPM no longer operates as a photo-detector for those illuminations. However, for maximum illumination, the SiPM retains a performance of 1.6. Nevertheless, this performance parameter is much lower than the performance value of 323 measured for the same LED illumination on Earth (at $-3$~$^{\mathrm{o}}$C).}
    \label{fig:perf_vs_time}
\end{figure}

In addition, the operation performance of SiPMs could be assessed for a given LED illumination by observing the ratio between the SiPM current at that illumination and the SiPM dark current. This performance parameter can be calculated for any LED light intensity in a standard test. When this parameter approaches a value of 1, it means that no signal can be detected from the SiPMs for a given LED illumination (i.e. the I-I curve is constant). The performance parameter vs. days since launch for various LED illuminations can be seen in Figure~\ref{fig:perf_vs_time}. 

The decrease in the performance parameter is mainly due to the increase of SiPM dark current, which turns comparable with generated photocurrent for those LED illuminations. For the maximum illumination, the SiPM retains a performance of 1.6. For 85 \% illumination, performance is very close to 1. Nevertheless, a performance parameter of 1.6 for maximum illumination is much lower than the performance of 323 measured (at $-3$~$^{\mathrm{o}}$C) for the same LED illumination on Earth.



Similar SiPM devices have been tested in radiation facilities by other authors. Table~\ref{tab:sipm_comp} shows SiPM model and radiation source for each work.

\begin{table}[!h]
    \centering
    \setlength{\tabcolsep}{3pt}
    \begin{tabular}{|c|c|c|}
    \hline
    \begin{tabular}{@{}c@{}} SiPM Model \end{tabular} & Source & Ref. \\
    \hline
    SensL MicroFC-60035 & 8~MeV electrons & \cite{sipm_comp1} \\
    SensL MicroFC-60035 & 64~MeV protons & \cite{sipm_comp1} \\
    SensL MicroFC-10050 & White neutrons & \cite{sipm_comp2, sipm_comp22} \\
    FBKVUV-HD 2019 & 75~MeV protons & \cite{sipm_comp3} \\
    \hline
    \end{tabular}
    \caption{SiPM models and radiation sources of several references that can be used to compare to this work.}
    \label{tab:sipm_comp}
\end{table}

\begin{figure}[!h]
\centering
    \includegraphics[width=0.7\textwidth]{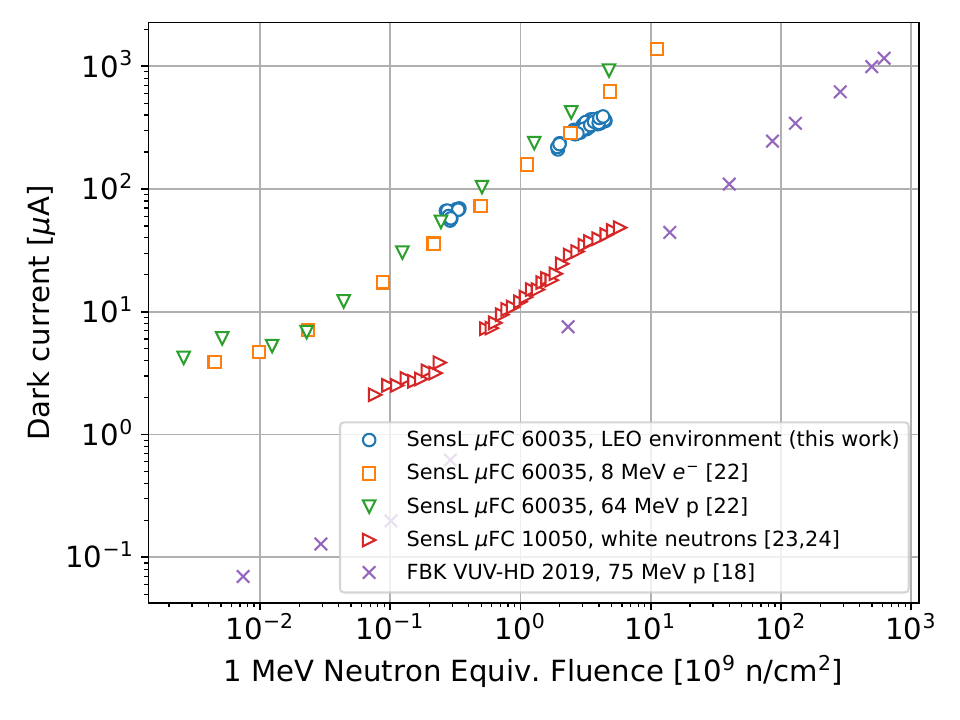}
    \caption{Comparison of dark current vs. 1~MeV neutron equivalent fluence between our measurements and cited references in Table~\ref{tab:sipm_comp}. SiPMs from those references are from different SiPM models and technologies and were irradiated with various radiation sources. It is interesting to observe that the increase in dark current with dose is the same, or very similar, for every case, irrespective of the initial (or non-irradiated) dark current. One of the cited references irradiates the same SiPM sensor we used (SensL MicroFC-60035) at similar overvoltage conditions and the dark current comparison is similar for a wide range of fluences.}
    \label{fig:damage_comparison}
\end{figure}

From each of these articles, measurements of dark current vs. 1~MeV equivalent fluence were extracted and can be compared to the measurements in this work. This comparison plot can be seen in Figure~\ref{fig:damage_comparison}.

It is interesting to note that all SiPMs present the same (or very similar) increase in dark current vs. dose, irrespective of their initial (or non-irradiated) dark current. SiPM models such as SensL MicroFC-10050 (1~mm$^2$) present a smaller dark current because of their reduced active area when compared to SensL MicroFC-60035 (36~mm$^2$). Our measurements seem to validate the 1~MeV equivalence estimation and the NIEL scaling hypothesis holds true for the studied SiPMs as well. This means that damage from LEO environments such as the one presented in this work can be correctly estimated by SiPM irradiation on Earth with different particles and different energies, so as long as the neutron equivalent fluence expected in orbit is the same. 

In addition, the dark current comparison between our measurements and MicroFC-60035 irradiated with protons and electrons~\cite{sipm_comp1} is strikingly similar. This is beside the fact that the LEO irradiation conditions differ greatly from a mono energetic proton or electron beam. In space, SiPMs are continuously being irradiated by a distribution of protons and electrons of different energies while, most of the time, they are at sub 0~$^{\mathrm{o}}$C temperatures. Moreover, they receive radiation in small doses over a prolonged time, as opposed to a large dose received at once.

\subsection{Long Test}

\begin{figure}[H]
\centering
    \includegraphics[width=0.65\textwidth]{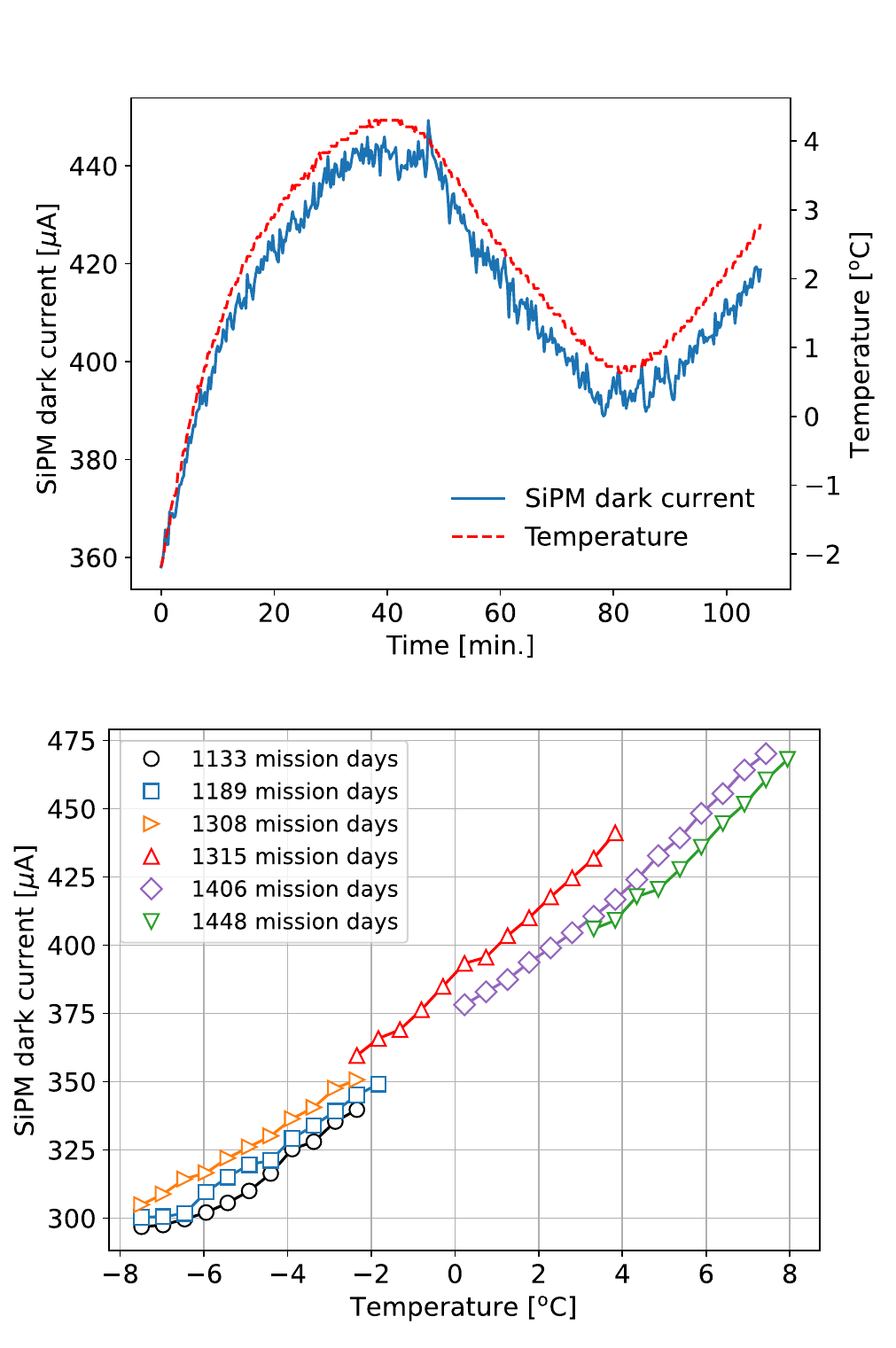}
    \caption{\textbf{Top plot:} Plot of SiPM current vs. time along with a plot of Temperature vs. orbit time. This measurement corresponds to 1180 days after launch (August 2023). \textbf{Bottom plot:} SiPM current vs. Temperature for several long tests performed in different days. A positive offset is observed in the curves with increasing time. This tendency was observed up to 1315 mission days, which is around the date of the last report before the satellite's pronounced loss in altitude. After that point, the tendency is reversed and one hypothesis is that the SiPMs are slightly recovering due to annealing from the increase in temperature.}
    \label{fig:sipm_temp}
\end{figure}

As previously stated, this test allowed for the measurement of SiPM dark current along a full satellite orbit. An example of a long test measurement for a particular SiPM can be seen in the top plot of Figure~\ref{fig:sipm_temp}.

An increase in SiPM temperature is observed in the first 40 minutes of the measurement, which is due to payload self heating at the start of the long test. Afterwards, the temperature oscillates due to the satellite being in Eclipse or in-Sun during its orbit. The correlation between SiPM dark current and temperature is also shown in the bottom plot of Figure~\ref{fig:sipm_temp}. This correlation shows a positive offset with increasing time. This tendency was observed up to 1315 mission days, which is around the date of the last report before the satellite's pronounced loss in altitude. After that point, the tendency is reversed. To our knowledge, the only experimental parameter that changed after that date was the satellite altitude, which is highly correlated to DB/SiPM temperature, as shown in Figure~\ref{fig:temp_vs_dsl}. Due to this, a very likely hypothesis is that the SiPM radiation damage was partially recovered due to annealing caused by the higher temperatures the SiPMs were subjected to. This behaviour is consistent with previous works~\cite{current_vs_temp_annealing}, where it is shown that the characteristic time of SiPM recovery due to annealing changes noticeably even for small temperature differences. Nevertheless, it is important to note that the recovery observed is very small and in no way approached SiPM dark currents before launch.

Another interesting result seen in the bottom plot of Figure~\ref{fig:temp_vs_dsl} is the reduction in SiPM dark current increase with temperature. Before launch, dark current doubled every 10~$^{\mathrm{o}}$C and this is no longer true after 1460 mission days. At the end of the mission, dark current was estimated to double every 20~$^{\mathrm{o}}$C instead. This reduction in SiPM dark current increase with temperature was studied in~\cite{current_vs_temp_annealing}, where SiPMs were irradiated with neutrons, and measurements obtained there are consistent with this work. 

\section{Conclusions and Outlook} \label{sec:conclusions}

In this work, the performance of SensL MicroFC-60035 SiPM devices was studied in Low Earth Orbit using the LabOSat-01 characterization payload. Four SiPMs were integrated into the ÑuSat-7 satellite (COSPAR-ID: 2020-003B) and the mission had a total duration of 1460 days. Throughout the mission, the SiPMs received damage from trapped and solar protons in the satellite's orbit. Different measurements were performed on the SiPMs to gain further insight on this damage and the SiPMs' operation characteristics. 

The dark current was observed to increase by a factor of 500 during the mission, while other characteristics, like Gain and Photon Detection Efficiency remained unchanged. It was also observed that SiPMs remained operational for the brighter LED light intensities under study, albeit with a reduced performance. The 1~MeV neutron equivalent fluence that the electronics and SiPMs received was measured to be $\phi_n = 5 \cdot 10^9$~n/cm$^2$ with p-MOSFET dosimeters integrated near our payload. Dark current increase for a given measured dose was compared with other works, with compatible results. Our measurements seem to validate the 1~MeV equivalence estimation and the NIEL scaling hypothesis holds true for the studied SiPMs as well. This means that damage from LEO environments such as the one presented in this work can be correctly estimated by SiPM irradiation on Earth with different particles and different energies, so as long as the neutron equivalent fluence expected in orbit is the same.

In addition, SiPMs were studied along full satellite orbits and interesting results were obtained. It was observed that the rate of increase of dark current with temperature is smaller after SiPMs are damaged by radiation.

\section*{Acknowledgements}

The authors would like to thank Satellogic for their help during the AIT, mission commissioning and operations. The authors acknowledge financial support from ANPCyT, PICT 2017-0984 ``Componentes Electrónicos para Aplicaciones Satelitales (CEpAS)'', PICT-2018-0365 ``LabOSat: Plataforma de caracterización de dispositivos electrónicos en ambientes hostiles'', PICT-2019-2019-02993 ``LabOSat: desarrollo de un Instrumento detector de fotones individuales para aplicaciones espaciales'' and UNSAM-ECyT FP-001.


\printbibliography

\end{document}